# Performance of Opportunistic Fixed Gain Bidirectional Relaying With Outdated CSI[†]

Fahd Ahmed Khan, *Student Member, IEEE*, Kamel Tourki, *Member, IEEE*, Mohamed-Slim Alouini, *Fellow, IEEE*, and Khalid A. Qaraqe, *Senior Member, IEEE*.

## Abstract

This paper studies the impact of using outdated channel state information for relay selection on the performance of a network where two sources communicate with each other via fixed-gain amplify-and-forward relays. For a Rayleigh faded channel, closed-form expressions for the outage probability, moment generating function and symbol error rate are derived. Simulations results are also presented to corroborate the derived analytical results. It is shown that adding relays does not improve the performance if the channel is substantially outdated. Furthermore, relay location is also taken into consideration and it is shown that the performance can be improved by placing the relay closer to the source whose channel is more outdated.

[†] This work was supported in part by King Abdullah University of Science and Technology (KAUST) and in part by NPRP grant number NPRP 5-250-2-087 from the Qatar National Research Fund (a member of Qatar Foundation). The statements made herein are solely the responsibility of the authors. Fahd Ahmed Khan and Mohamed-Slim Alouini are with the Computer, Electrical, and Mathematical Sciences and Engineering (CEMSE) Division, KAUST, Thuwal, Makkah Province, Saudi Arabia. Email: {fahd.khan,slim.alouini}@kaust.edu.sa. Kamel Tourki and Khalid A. Qaraqe are with the ECEN Program, Texas A&M Univ. at Qatar, Doha, Qatar. Email: {kamel.tourki,khalid.qaraqe}@qatar.tamu.edu. The corresponding author is Fahd Ahmed Khan (fahd.khan@kaust.edu.sa).

# I. Introduction

Cooperative relaying helps mitigate the fading and achieve spatial diversity gains. However, a drawback of employing relays is the loss of throughput/spectral-efficiency due to orthogonal signalling [1]. Two-way relaying (or Bidirectional relaying), where two users exchange information via common intermediate relays, has been proposed to improve the throughput compared to traditional one-way relaying while maintaining its diversity benefits [2]–[5].

The throughput of cooperative relay networks can be further improved by employing relay selection (RS) where a single relay is opportunistically selected for transmission and thus, the throughput is improved compared to the system where all the relays participate. It has also been shown that RS also preserves the same spatial diversity [6]–[8]. Many works have analysed the performance of relay selection in context of two-way relaying eg. see [9]–[15] and references therein. In [9],the performance of the max-min RS technique and the max-sum RS technique was analysed and a hybrid RS scheme was proposed. It was shown that max-min RS scheme extracts the full diversity gain and is suitable at high signal-to-noise ratio (SNR). Performance of various full diversity achieving schemes based on joint network coding and opportunistic relaying have been analysed in [10] and [11]. For bidirectional amplify-and-forward (AF) relaying, the performance of opportunistic RS based on the max-min criteria was analysed in terms of outage probability and symbol error rate (SER) in [12]–[15] and it was shown that this scheme achieves full diversity.

All these previous works assumed that the channel remains the same during the RS phase and the data transmission phase. This assumption is not always true and it is possible that due to feedback delay or scheduling delay, an outdated channel is used for RS. This impacts the performance of the system significantly and for traditional one-way relay networks, it is shown that RS using outdated CSI (OC) results in diversity loss eg. see [16]–[20]. This loss in performance and diversity also occurs for an opportunistic two-way relaying network (OTN) with OC based RS [21], [22]. The performance of an OTN with max-min RS (MRS) scheme

                                                                    

based on OC was analysed in [21] and the lower and upper bounds on the outage performance and SER performance were derived. It was also shown that the diversity is lost due to OC. The authors in [22] also considered an OTN with max-min RS and OC (OTN-MRS-OC), and obtained closed-form expressions for outage probability and SER.

The authors in [21] and [22] considered variable gain AF relays. In this paper, we analyse the performance of OTN-MRS-OC with fixed gain (FG) AF relays. The advantage of employing FG relays is that the relays have lower complexity and thus, less cost, because the channel is not estimated at the relay. Instead the relay only requires the long term statistics of the channel which vary slowly and can thus be easily estimated without additional complexity/overhead [23]. Furthermore, for FG relays, the statistics of the end-to-end SNR (E2E-SNR) are different from the ones obtained for variable gain relays in [21] and [22]. To the best of the authors knowledge the performance of OTN-MRS-OC with FG relays has not analysed previously. Thus, in this paper we obtain the closed-form expressions for the outage performance, moment-generating-function of the E2E-SNR and the SER performance for both coherent and non-coherent one-dimensional modulation schemes. Numerical simulation results are also presented to validate the derived analytical results. In addition, unlike previous works considering RS based on OC, the impact of relay location is also studied and it is shown that the performance of the OTN-MRS-OC can be improved by moving the relay closer to the source whose channel is more outdated.

## II. System Model

Consider a system where two sources, denoted by S1 and S2, communicate with each other via $R$ two-way AF relays. The direct link between both the sources is assumed to be deeply faded. A single relay is opportunistically selected for transmission and data transmission occurs after the relay selection. Data transmission in a two-way relaying network is carried out in two phases (i) multiple access phase and (ii) broadcast phase. In the multiple access phase both the sources transmit simultaneously to the relay where as in the broadcast phase the relay transmits to both the sources.

We consider a realistic scenario, where there is a delay between the RS phase and data transmission. Due to this delay, the channel becomes outdated i.e. the channel conditions during the data transmission are different from those during the RS phase. During the RS phase (past realizations), $\tilde{g}_{1,n}$ and $\tilde{g}_{2,n}$ are used to denote the channel gains, between S1 and the $n$-th relay and between S2 and the $n$-th relay, respectively. Similarly, $g_{1,n}$ and $g_{2,n}$ are the channel gains during the data transmission (recent realizations). The channel is assumed to be independent and identically distributed (i.i.d) and have Rayleigh fading.

For a Rayleigh fading channel, the past and current realizations of the channel gains are related with each other according to

$$g_{1,n} = \rho_1 \tilde{g}_{1,n} + \sqrt{1 - \rho_1^2} v_n, \quad g_{2,n} = \rho_2 \tilde{g}_{2,n} + \sqrt{1 - \rho_2^2} w_n, \tag{1}$$

where $\rho_1 = J_0 \left(2\pi f_{D,1} T\right)$ and $\rho_2 = J_0 \left(2\pi f_{D,2} T\right)$ are the correlation coefficients, $T$ denotes the time delay between the RS phase and the data transmission, $f_{D,1}$ is the maximum Doppler frequency shift of the channel between S1 and the relays, $f_{D,2}$ is the maximum Doppler frequency shift of the channel between the relays and S2, $J_0(\cdot)$ denotes the zero-order Bessel function of the first kind. $\tilde{g}_{1,n}$ and $v_n$ are both zero mean complex Gaussian (ZMCG) random variables (RVs) with variance $\sigma_1$. Similarly, $\tilde{g}_{2,n}$ and $w_n$ are both ZMCG RVs with variance $\sigma_2$.

Given that the $n$-th relay is selected for transmission, the end-to-end SNR (E2E-SNR), after removing the self interference, at S2 can be expressed as

$$\Upsilon_n = \frac{P_S G^2 \left|g_{2,n}\right|^2 \left|g_{1,n}\right|^2}{G^2 \left|g_{2,n}\right|^2 N_0 + N_0} = \frac{P_S \left|g_{2,n}\right|^2 \left|g_{1,n}\right|^2}{\left|g_{2,n}\right|^2 N_0 + \frac{N_0}{G^2}} \tag{2}$$

where $P_S$ is the transmit power of the sources, $P_R$ is the transmit power of the relays, $N_0$ is the variance of the additive ZMCG noise at the sources and the relays, and $G$ denotes the amplification gain at the relays. For fixed gain relays, $G = \sqrt{\frac{P_R}{P_S \mathbf{E}\left[|\tilde{g}_{1,n}|^2\right] + P_S \mathbf{E}\left[|\tilde{g}_{2,n}|^2\right] + N_0}}$, therefore the end-to-end SNR from S1 to S2 can be expressed as

$$\Upsilon_n = \frac{\eta_1 \eta_2 \gamma_{1,n} \gamma_{2,n}}{\eta_2 \gamma_{2,n} + \overline{C}} \tag{3}$$

where $\gamma_{1,n} = |g_{1,n}|^2$, $\gamma_{2,n} = |g_{2,n}|^2$, $\eta_1 = \frac{P_S}{N_0}$, $\eta_2 = \frac{P_R}{N_0}$, $\bar{C} = \left( \eta_1 \left( \mathbf{E}\left[|\tilde{g}_{1,n}|^2\right] + \mathbf{E}\left[|\tilde{g}_{2,n}|^2\right] \right) + 1 \right)$ and $\mathbf{E}\left[\cdot\right]$ denotes the expectation operator. Note that $\tilde{\gamma}_{1,n} = |\tilde{g}_{1,n}|^2$ and $\tilde{\gamma}_{2,n} = |\tilde{g}_{2,n}|^2$ are the corresponding channel power gains during the RS phase. For an i.i.d Rayleigh fading channel, $\tilde{\gamma}_{1,n}$ and $\tilde{\gamma}_{2,n}$ are i.i.d exponential RVs with mean $\sigma_1$ and $\sigma_2$, respectively, $\forall\, n$. The PDF and CDF of an exponential RV $X$ with mean $\mu$ is given as $f_X(x) = \frac{1}{\mu} e^{-\frac{x}{\mu}}$ and $F_X(x) = \left(1 - e^{-\frac{x}{\mu}}\right)$, respectively. Furthermore, following (1), it can be shown that $\tilde{\gamma}_{i,n}$ and $\gamma_{i,n}$ are two correlated exponential RVs and their joint PDF is given as [16]

$$f_{\tilde{\gamma}_{i,n}, \gamma_{i,n}}(y,x) = \frac{1}{(1-\rho_i^2)\sigma_i^2} e^{-\frac{x+y}{(1-\rho_i^2)\sigma_i}} I_0 \left( \frac{2\sqrt{\rho_i^2\, x\, y}}{(1-\rho_i^2)\, \sigma_i} \right),  \tag{4}$$

where, $i = 1$ denotes the S1 to relay link, $i = 2$ denotes the S2 to relay link and $\sigma_i$ is the mean power of the source-$i$ to relay link.

For opportunistic relaying, the E2E-SNR can be expressed as

$$\Upsilon_{\mathcal{F}} = \frac{\eta_1 \eta_2 \gamma_{1,eq} \gamma_{2,eq}}{\eta_2 \gamma_{2,eq} + C}  \tag{5}$$

where $C = (\eta_1 (\mathbf{E}\left[\tilde{\gamma}_{1,eq}\right] + \mathbf{E}\left[\tilde{\gamma}_{2,eq}\right]) + 1)$, $\gamma_{1,eq}$ is the effective instantaneous channel power gain of the S1 to relay link and $\gamma_{2,eq}$ is the effective instantaneous channel power gain of the S2 to relay link[1]. Both $\gamma_{1,eq}$ and $\gamma_{2,eq}$ depends on the RS criteria. In this paper, the RS is done based on the Max-Min criteria where the selected relay is $k = \arg\max_i \{\min\{\tilde{\gamma}_{1,i}, \tilde{\gamma}_{2,i}\}\} = \arg\max_i \{\Lambda_i\}$. It can be noted that the RS criteria uses the past realization of the channel power gain.

## III. Performance Analysis

In order to analyse the performance of the OTWRN-OC, the statistic (CDF and PDF) of $\Upsilon_{\mathcal{F}}$ are required.

---

[1] Without loss of generality, similar to (5), the E2E-SNR at S1 can be obtained by interchanging the indices 1 and 2. In this sequel, we present the performance analysis based on the E2E-SNR at S2. The performance at S1 can be obtained by interchanging the indices 1 and 2.

*A. CDF and PDF of $\Upsilon_{\mathcal{F}}$*

As $\Upsilon_{\mathcal{F}}$ depends on recent realizations $\gamma_{1,eq}$ and $\gamma_{2,eq}$ which are correlated with their past realization $\tilde{\gamma}_{1,eq}$ and $\tilde{\gamma}_{2,eq}$, therefore, first the CDF and PDF of $\tilde{\gamma}_{1,eq}$ and $\tilde{\gamma}_{2,eq}$ are derived. Following the steps given in Appendix A, the CDF and the PDF of $\tilde{\gamma}_{1,eq}$ and $\tilde{\gamma}_{2,eq}$ can be obtained. Using the derived PDFs and CDFs in (19), (20), (21) and (22), and following the procedure in Appendix B, the PDFs and CDFs of the equivalent channel power gain during the data transmission phase ($\gamma_{1,eq}$ and $\gamma_{2,eq}$) are obtained in (28), (29), (30) and (31).

Using (5), the CDF of $\Upsilon_{\mathcal{F}}$ can be expressed as

$$F_{\Upsilon_{\mathcal{F}}}(\Phi) = \Pr\left\{\Upsilon_{\mathcal{F}} < \Phi\right\} = \Pr\left\{\frac{\eta_1\eta_2\gamma_{1,eq}\gamma_{2,eq}}{\eta_2\gamma_{2,eq}+C} < \Phi\right\} = \Pr\left\{\gamma_{1,eq} < \frac{\Phi\left(\eta_2\gamma_{2,eq}+C\right)}{\eta_1\eta_2\gamma_{2,eq}}\right\}, \quad (6)$$

which can be evaluated by conditioning on $\gamma_{2,eq}$ and then averaging using the PDF of $\gamma_{2,eq}$ as

$$F_{\Upsilon_{\mathcal{F}}}(\Phi) = \int_0^\infty F_{\gamma_{1,eq}}\left(\frac{\Phi\left(\eta_2\gamma+C\right)}{\eta_1\eta_2\gamma}\right)f_{\gamma_{2,eq}}(\gamma)\,d\gamma. \quad (7)$$

Substituting the CDF from (30) and PDF from (29) into (7) and doing some algebraic manipulations one gets

$$F_{\Upsilon_{\mathcal{F}}}(\Phi) = R^2\sum_{i_1=0}^{R-1}\sum_{j_1=2}^{3}\sum_{i_2=0}^{R-1}\sum_{j_2=2}^{3}\binom{R-1}{i_1}\binom{R-1}{i_2}(-1)^{i_1+i_2}\alpha_{1,j_1,i_1}\alpha_{2,j_2,i_2}$$
$$\left(1 - \frac{\nu_2\beta_{2,j_2,i_2}}{(\beta_{2,j_2,i_2}+\rho_2^2\nu_2)}e^{-\frac{\nu_1\beta_{1,j_1,i_1}}{(\beta_{1,j_1,i_1}+\rho_1^2\nu_1)}\frac{\Phi}{\eta_1}}\int_0^\infty e^{-\frac{\nu_2\beta_{2,j_2,i_2}}{(\beta_{2,j_2,i_2}+\rho_2^2\nu_2)}\gamma-\frac{\nu_1\beta_{1,j_1,i_1}}{(\beta_{1,j_1,i_1}+\rho_1^2\nu_1)}\frac{\Phi C}{\eta_1\eta_2\gamma}}d\gamma\right) \quad (8)$$

Solving the integration using [24, Eq. (3.478.4)], and doing some algebraic manipulations yields

$$F_{\Upsilon_{\mathcal{F}}}(\Phi) = \sum_{\mathcal{F}}\mathcal{X}_{\mathcal{F}}\left(1 - 2e^{-\Theta_1\frac{\Phi}{\eta_1}}\sqrt{\frac{\Theta_1\Theta_2 C}{\eta_1\eta_2}\Phi}K_1\left(2\sqrt{\frac{\Theta_1\Theta_2 C}{\eta_1\eta_2}\Phi}\right)\right), \quad (9)$$

where $\Theta_1 = \frac{\beta_{1,j_1,i_1}}{(\bar{v}_1\beta_{1,j_1,i_1}+\rho_1^2)}$, $\Theta_2 = \frac{\beta_{2,j_2,i_2}}{(\bar{v}_2\beta_{2,j_2,i_2}+\rho_2^2)}$, $\bar{v}_1 = (1-\rho_1^2)\sigma_1$, $\bar{v}_2 = (1-\rho_2^2)\sigma_2$, $\mathcal{X}_{\mathcal{F}} = R^2\binom{R-1}{i_1}\binom{R-1}{i_2}(-1)^{i_1+i_2}\alpha_{1,j_1,i_1}\alpha_{2,j_2,i_2}$ and $\sum_{\mathcal{F}}$ is shorthand notation of $\sum_{i_1=0}^{R-1}\sum_{j_1=2}^{3}\sum_{i_2=0}^{R-1}\sum_{j_2=2}^{3}$. The PDF of $\Upsilon_{\mathcal{F}}$ can be obtained by taking the derivative of the CDF in (9). Using the CDF in (9), various performance metric such as the MGF, outage probability, and SER can be derived.

## B. Outage Probability

Using the CDF in (9), the outage performance of the OTWRN-OC can be obtained as

$$\mathcal{O}\left(\Psi\right) = F_{\Upsilon_{\mathcal{F}}}\left(\Psi\right) \tag{10}$$

where $\Psi = 2^{\mathcal{R}} - 1$ and $\mathcal{R}$ is the transmission rate.

## C. MGF and Symbol Error Rate Performance

The MGF can be obtained using the CDF of the E2E-SNR as [25, Eq. (18)]

$$\mathcal{M}_{\Upsilon_{\mathcal{F}}}(s) = s \int_0^\infty e^{-sx} F_{\Upsilon_{\mathcal{F}}}(x) dx \tag{11}$$

The SER for non-coherent modulation schemes can be obtained using the CDF of the E2E-SNR as

$$\mathcal{P}_{e,\mathcal{NC}} = a\,b \int_0^\infty e^{-bx} F_{\Upsilon_{\mathcal{F}}}(x) dx \tag{12}$$

where $a$ and $b$ are modulation-specific constants eg. $(a,b) = (0.5, 1)$ for DBPSK and $(a,b) = (0.5, 0.5)$ for NCBFSK [26]. The SER for coherent modulation schemes can be obtained using the CDF of the E2E-SNR as

$$\mathcal{P}_{e,\mathcal{C}} = \frac{a}{2}\sqrt{\frac{b}{2\pi}} \int_0^\infty x^{-\frac{1}{2}} e^{-\frac{b}{2}x} F_{\Upsilon_{\mathcal{F}}}(x) dx \tag{13}$$

where $(a,b) = (1,2)$ for BPSK, $(a,b) = (1,1)$ for BFSK, $(a,b) = \left(2\frac{M-1}{M}, 6\frac{\log_2(M)}{M^2-1}\right)$ for $M$-PAM [26].

The integrals in (11), (12) and (13) are of type

$$S(c_1, c_2, c_3) = c_1 \int_0^\infty x^{c_2} e^{-c_3 x} F_{\Upsilon_{\mathcal{F}}}(x) dx. \tag{14}$$

and thus, the MGF and the SER can be expressed in terms of $S(\cdot, \cdot, \cdot)$ i.e. $\mathcal{M}_{\Upsilon_{\mathcal{F}}}(s) = \mathcal{S}\left(s, 0, s\right)$, $\mathcal{P}_{e,\mathcal{NC}} = \mathcal{S}\left(a\,b, 0, b\right)$ and $\mathcal{P}_{e,\mathcal{C}} = \mathcal{S}\left(\frac{a}{2}\sqrt{\frac{b}{2\pi}}, -\frac{1}{2}, \frac{b}{2}\right)$. The closed-form solution of $S(\cdot, \cdot, \cdot)$ is derived in Appendix-C. The solution in (35) involves a Gamma function and a Meijer-G function which are available in well known mathematical packages and thus, the SER performance can be easily and accurately evaluated.

## IV. Numerical Results and Discussion

In this section, some selected numerical results as well as Monte-Carlo based simulation results are presented to verify the derived analytical results. In obtaining these numerical results, $\mathcal{R} = 1$, $N_0 = 1$, $\eta_1 = \eta_2$. These parameters are fixed in the simulation unless stated. The effect of path-loss is captured by taking $\sigma_1 = d_1^{-\upsilon}$ and $\sigma_2 = d_2^{-\upsilon} = (1 - d_1)^{-\upsilon}$, where $d_i$ is the distance of source-$i$ from the relays and $\upsilon$ is the path-loss exponent. Note that the distances $d_i$ are normalized w.r.t. the distance between both sources.

Fig. 1 shows the effect of varying the relay location on the outage probability performance of the . It can be observed that the performance degrades as the correlation, $\rho_i$, reduces. For the case when $\rho_1 = \rho_2 = 1$, the best performance is achieved when $d_1 = 0.5$, i.e. the relay is in the middle of both sources. If $\rho_1 < \rho_2$ or $\rho_1 > \rho_2$, $d_1 = 0.5$ does not give best outage performance. For $\rho_1 < \rho_2$, the outage probability can be lowered by reducing $d_1$ and vice-verse. Furthermore, it can be observed that increasing the number of relays improves the outage performance only if the correlation is sufficiently high. If the correlation is very low then, adding relays has no benefit as can be observed for the case when $\rho_1 = \rho_2 = 0.2$.

Fig. 2 shows the SER of BPSK modulation scheme as a function of $\eta_1$ when the relays are in the middle of the sources i.e. ($d_1 = 0.5$). It can be observed that when $\rho_i < 1$, the performance degrades severely and diversity is lost. Furthermore, SER increases as correlation decreases. Again it can be noticed that increasing the number of relays improves the performance only if the correlation is sufficiently high. When correlation is very less, adding relays can even degrade performance as can be observed for $\rho_1 = \rho_2 = 0.2$. The SER performance however, can be improved by varying $d_1$ and finding the optimal relay position. Note that, in both figures, the simulation results match well with the analytical results.

## V. Conclusion

In this work, the performance of max-min relay selection based on outdated CSI in a two-way relay network is analysed. The relays are assumed to be fixed gain amplify-and-forward relays.

Expressions for the outage probability, moment generating function and symbol error rate are derived for a Rayleigh faded channel. These expressions are validated by numerical simulation. Numerical simulation results show that the diversity is lost due to OC. Furthermore, relay location is also taken into consideration and it is shown that the performance can be improved by placing the relay closer to the source whose channel is more outdated.

## VI. APPENDIX

### A. CDF and PDF of $\tilde{\gamma}_{1,eq}$ and $\tilde{\gamma}_{2,eq}$

The CDF of $\tilde{\gamma}_{1,eq}$ can be obtained as

$$F_{\tilde{\gamma}_{1,eq}}(\Phi) = R\left(\Pr\left(\tilde{\gamma}_{1,n} < \Phi \bigcap \tilde{\gamma}_{1,n} > \tilde{\gamma}_{2,n} \bigcap n = k\right) + \Pr\left(\tilde{\gamma}_{1,n} < \Phi \bigcap \tilde{\gamma}_{1,n} < \tilde{\gamma}_{2,n} \bigcap n = k\right)\right). \quad (15)$$

Note that $n = k$ denotes that relay $k$ is selected for transmission. $F_{\tilde{\gamma}_{1,eq}}(\cdot)$ can be expressed as

$$F_{\tilde{\gamma}_{1,eq}}(\Phi) = R\left(\int_0^\Phi f_{\tilde{\gamma}_{1,n}}(x) \int_0^x f_{\tilde{\gamma}_{2,n}}(y) \prod_{i\neq n} F_{\Lambda_i}(x)\, dy dx + \int_0^\Phi f_{\tilde{\gamma}_{1,n}}(x) \prod_{i\neq n} F_{\Lambda_i}(x) \int_x^\infty f_{\tilde{\gamma}_{2,n}}(y)\, dy dx\right) \quad (16)$$

where $f_{\tilde{\gamma}_{1,n}}(\cdot)$ and $f_{\tilde{\gamma}_{2,n}}(\cdot)$ denote the PDF of $\tilde{\gamma}_{1,n}$ and $\tilde{\gamma}_{2,n}$, respectively and $F_{\Lambda_i}(\cdot)$ denotes the CDF of $\Lambda_i$. Substituting the PDF and CDF one gets

$$
\begin{aligned}
F_{\tilde{\gamma}_{1,eq}}(\Phi) = R\Bigg(&\int_0^\Phi \frac{1}{\sigma_1} e^{-\frac{x}{\sigma_1}} \int_0^x \frac{1}{\sigma_2} e^{-\frac{y}{\sigma_2}} \left(1 - e^{-\left(\frac{1}{\sigma_1}+\frac{1}{\sigma_2}\right)y}\right)^{R-1} dy dx \\
&+ \int_0^\Phi \frac{1}{\sigma_1} e^{-\frac{x}{\sigma_1}} \left(1 - e^{-\left(\frac{1}{\sigma_1}+\frac{1}{\sigma_2}\right)x}\right)^{R-1} \int_x^\infty \frac{1}{\sigma_2} e^{-\frac{y}{\sigma_2}} dy dx\Bigg)
\end{aligned}
\quad (17)
$$

where $F_{\Lambda_i}(x) = 1 - e^{-\left(\frac{1}{\sigma_1}+\frac{1}{\sigma_2}\right)x}$. Using binomial expansion, integrating w.r.t $y$ and $x$ and doing some algebraic manipulations yields

$$F_{\tilde{\gamma}_{1,eq}}(\Phi) = R \sum_{i=0}^{R-1} \binom{R-1}{i} (-1)^i \left(\frac{\left(1 - e^{-\frac{\Phi}{\sigma_1}}\right)}{\sigma_2\left(\chi_i - \frac{1}{\sigma_1}\right)} - \left(\frac{1}{\sigma_1\sigma_2\left(\chi_i - \frac{1}{\sigma_1}\right)} - \frac{1}{\sigma_1}\right) \frac{1}{\chi_i}\left(1 - e^{-\chi_i\Phi}\right)\right) \quad (18)$$

where $\chi_i = \left(\frac{i+1}{\sigma_1} + \frac{i+1}{\sigma_2}\right)$. $F_{\tilde{\gamma}_{1,eq}}(\Phi)$ can be compactly expressed as

$$F_{\tilde{\gamma}_{1,eq}}(\Phi) = R \sum_{i=0}^{R-1} \sum_{j=1}^3 \binom{R-1}{i} (-1)^i \alpha_{1,j,i} e^{-\beta_{1,j,i}\Phi} \quad (19)$$

where $\alpha_{1,1,i} = (\kappa_{1,1,i} - \kappa_{1,2,i})$, $\alpha_{1,2,i} = -\kappa_{1,1,i}$, $\alpha_{1,3,i} = \kappa_{1,2,i}$, $\beta_{1,1,i} = 0$, $\beta_{1,2,i} = \frac{1}{\sigma_1}$, $\beta_{1,3,i} = \chi_i$, $\kappa_{1,1,i} = \frac{1}{\sigma_2\left(\chi_i - \frac{1}{\sigma_1}\right)}$ and $\kappa_{1,2,i} = \left(\frac{1}{\sigma_1\sigma_2\left(\chi_i - \frac{1}{\sigma_1}\right)\chi_i} - \frac{1}{\sigma_1\chi_i}\right)$. Similarly the expression for $F_{\tilde{\gamma}_{2,eq}}(\cdot)$ can be obtained by interchanging indices 1 and 2 yielding

$$F_{\tilde{\gamma}_{2,eq}}(\Phi) = R \sum_{i=0}^{R-1} \sum_{j=1}^{3} \binom{R-1}{i} (-1)^i \alpha_{2,j,i} e^{-\beta_{2,j,i}\Phi} \tag{20}$$

where $\alpha_{2,1,i} = (\kappa_{2,1,i} - \kappa_{2,2,i})$, $\alpha_{2,2,i} = -\kappa_{2,1,i}$, $\alpha_{2,3,i} = \kappa_{2,2,i}$, $\beta_{2,1,i} = 0$, $\beta_{2,2,i} = \frac{1}{\sigma_2}$, $\beta_{2,3,i} = \chi_i$, $\kappa_{2,1,i} = \frac{1}{\sigma_1\left(\chi_i - \frac{1}{\sigma_2}\right)}$ and $\kappa_{2,2,i} = \left(\frac{1}{\sigma_1\sigma_2\left(\chi_i - \frac{1}{\sigma_2}\right)\chi_i} - \frac{1}{\sigma_2\chi_i}\right)$.

The PDF of $\tilde{\gamma}_{1,eq}$ is obtained by taking the derivative of the CDF in (19) to give

$$f_{\tilde{\gamma}_{1,eq}}(\Phi) = R \sum_{i=0}^{R-1} \sum_{j=2}^{3} \binom{R-1}{i} (-1)^{i+1} \alpha_{1,j,i} \beta_{1,j,i} e^{-\beta_{1,j,i}\Phi}, \tag{21}$$

and the PDF of $\tilde{\gamma}_{2,eq}$ is obtained by taking the derivative of the CDF in (20) to give

$$f_{\tilde{\gamma}_{2,eq}}(\Phi) = R \sum_{i=0}^{R-1} \sum_{j=2}^{3} \binom{R-1}{i} (-1)^{i+1} \alpha_{2,j,i} \beta_{2,j,i} e^{-\beta_{2,j,i}\Phi}. \tag{22}$$

The mean of $\tilde{\gamma}_{q,eq}$, where $q \in \{1, 2\}$, is given as

$$\mathbf{E}[\tilde{\gamma}_{q,eq}] = R \sum_{i=0}^{R-1} \sum_{j=2}^{3} \binom{R-1}{i} (-1)^{i+1} \frac{\alpha_{q,j,i}}{\beta_{q,j,i}} \tag{23}$$

### B. CDF and PDF of $\gamma_{1,eq}$ and $\gamma_{2,eq}$

The PDF of $\gamma_{1,eq}$, $f_{\gamma_{1,eq}}(\cdot)$, can be obtained as

$$f_{\gamma_{1,eq}}(x) = \int_0^\infty f_{\tilde{\gamma}_{1,eq}, \gamma_{1,eq}}(y, x)\, dy \tag{24}$$

where $f_{\tilde{\gamma}_{1,eq}, \gamma_{1,eq}}(\cdot, \cdot)$ denotes the joint PDF of $\gamma_{1,eq}$ and $\tilde{\gamma}_{1,eq}$ and is given as

$$f_{\tilde{\gamma}_{1,eq}, \gamma_{1,eq}}(y, x) = \frac{f_{\tilde{\gamma}_{1,n}, \gamma_{1,n}}(y, x)}{f_{\tilde{\gamma}_{1,n}}(y)} f_{\tilde{\gamma}_{1,eq}}(y) \tag{25}$$

Substituting the PDF from (4) and (21) and into (25) and doing some algebraic manipulations yields

$$f_{\tilde{\gamma}_{1,eq}, \gamma_{1,eq}}(y, x) = R \sum_{i=0}^{R-1} \sum_{j=2}^{3} \binom{R-1}{i} (-1)^{i+1} \nu_1 \alpha_{1,j,i} \beta_{1,j,i} e^{-\nu_1 x - (\beta_{1,j,i} + \rho_1^2 \nu_1) y} I_0 \left(2\sqrt{\rho_1^2 \nu_1^2 x\, y}\right) \tag{26}$$

where $\nu_1 = \frac{1}{(1-\rho_1^2)\sigma_1}$. Similarly, the joint PDF of $\gamma_{2,eq}$ and $\tilde{\gamma}_{2,eq}$, $f_{\tilde{\gamma}_{2,eq}, \gamma_{2,eq}}(\cdot, \cdot)$ can be obtained as

$$f_{\tilde{\gamma}_{2,eq}, \gamma_{2,eq}}(y, x) = R \sum_{i=0}^{R-1} \sum_{j=2}^{3} \binom{R-1}{i} (-1)^{i+1} \nu_2 \alpha_{2,j,i} \beta_{2,j,i} e^{-\nu_2 x - (\beta_{2,j,i} + \rho_2^2 \nu_2) y} I_0 \left(2\sqrt{\rho_2^2 \nu_2^2 x\, y}\right) \tag{27}$$

where $\nu_2 = \frac{1}{(1-\rho_2^2)\sigma_2}$. Substituting the joint PDF from (26) into (24) and solving the resulting integration using [24], $\int_0^\infty e^{-a\,x} I_0 \left(2\,b\,\sqrt{x}\right) dx = \frac{1}{a} e^{\frac{b^2}{a}}$, $f_{\gamma_1,eq}\left(\cdot\right)$ is given as

$$f_{\gamma_1,eq}\left(x\right) = R \sum_{i=0}^{R-1} \sum_{j=2}^{3} \binom{R-1}{i} (-1)^{i+1} \frac{\nu_1 \alpha_{1,j,i} \beta_{1,j,i}}{(\beta_{1,j,i} + \rho_1^2 \nu_1)} e^{-\frac{\nu_1 \beta_{1,j,i}}{(\beta_{1,j,i} + \rho_1^2 \nu_1)}x} \tag{28}$$

Similarly, the PDF of $\gamma_{2,eq}$, $f_{\gamma_2,eq}\left(\cdot\right)$, is obtained by interchanging indices 1 and 2 yielding

$$f_{\gamma_2,eq}\left(x\right) = R \sum_{i=0}^{R-1} \sum_{j=2}^{3} \binom{R-1}{i} (-1)^{i+1} \frac{\nu_2 \alpha_{2,j,i} \beta_{2,j,i}}{(\beta_{2,j,i} + \rho_2^2 \nu_2)} e^{-\frac{\nu_2 \beta_{2,j,i}}{(\beta_{2,j,i} + \rho_2^2 \nu_2)}x} \tag{29}$$

The CDF of $\gamma_{1,eq}$, $F_{\gamma_1,eq}\left(\cdot\right)$, can be obtained by integrating $f_{\gamma_1,eq}\left(\cdot\right)$ to give

$$F_{\gamma_1,eq}\left(x\right) = R \sum_{i=0}^{R-1} \sum_{j=2}^{3} \binom{R-1}{i} (-1)^{i+1} \alpha_{1,j,i} \left(1 - e^{-\frac{\nu_1 \beta_{1,j,i}}{(\beta_{1,j,i} + \rho_1^2 \nu_1)}x}\right) \tag{30}$$

Similarly, the CDF of $\gamma_{2,eq}$, $F_{\gamma_2,eq}\left(\cdot\right)$, can be obtained by integrating $f_{\gamma_2,eq}\left(\cdot\right)$ to give

$$F_{\gamma_2,eq}\left(x\right) = R \sum_{i=0}^{R-1} \sum_{j=2}^{3} \binom{R-1}{i} (-1)^{i+1} \alpha_{2,j,i} \left(1 - e^{-\frac{\nu_2 \beta_{2,j,i}}{(\beta_{2,j,i} + \rho_2^2 \nu_2)}x}\right) \tag{31}$$

## C. Closed Form Solution of $S(c_1, c_2, c_3)$

$S(\cdot,\cdot,\cdot)$ is defined as

$$S(c_1, c_2, c_3) = c_1 \int_0^\infty x^{c_2} e^{-c_3 x} F_{\Upsilon_{\mathcal{F}}}(x) dx \tag{32}$$

Substituting the CDF from (9) into (32) and doing some algebraic manipulations yields

$$\mathcal{S}(c_1, c_2, c_3) = c_1 \sum_{\mathcal{F}} \mathcal{X}_{\mathcal{F}} \left( \int_0^\infty x^{c_2} e^{-c_3 x} dx - 2 \int_0^\infty x^{c_2} e^{-\left(c_3 + \frac{\Theta_1}{\eta_1}\right)x} \sqrt{\frac{\Theta_1 \Theta_2 C}{\eta_1 \eta_2}} x K_1 \left(2\sqrt{\frac{\Theta_1 \Theta_2 C}{\eta_1 \eta_2}} x\right) dx \right) \tag{33}$$

Representing $K_1\left(\cdot\right)$ in terms of Meijer-G function [27, Eq. (03.04.26.0008.01)] and applying the

scaling property yields

$$\mathcal{S}(c_1, c_2, c_3) = c_1 \sum_{\mathcal{F}} \mathcal{X}_{\mathcal{F}} \left( c_3^{-1-c_2} \Gamma\left(1 + c_2\right) - \left(\frac{\Theta_1 \Theta_2 C}{\eta_1 \eta_2}\right)^{-c_2} \int_0^\infty e^{-\left(c_3 + \frac{\Theta_1}{\eta_1}\right)x} G_{0,2}^{2,0} \left(\frac{\Theta_1 \Theta_2 C}{\eta_1 \eta_2} x \middle| \begin{array}{c} -,- \\ c_2 + 1, c_2 \end{array}\right) dx \right) \tag{34}$$

Solving the integration using [24, Eq. (7.813.1)] and applying the scaling property yields

$$\mathcal{S}(c_1, c_2, c_3) = c_1 \sum_{\mathcal{F}} \mathcal{X}_{\mathcal{F}} \left( c_3^{-1-c_2} \Gamma\left(1 + c_2\right) - \left(\frac{\Theta_1 \Theta_2 C}{\eta_1 \eta_2}\right)^{-c_2-1} G_{1,2}^{2,1} \left( \left(c_3 + \frac{\Theta_1}{\eta_1}\right)^{-1} \frac{\Theta_1 \Theta_2 C}{\eta_1 \eta_2} \middle| \begin{array}{c} 1,- \\ c_2 + 2, c_2 + 1 \end{array}\right) \right) \tag{35}$$

where $G_{p,q}^{m,n}\left(\cdot\right)$ is the Meijer-G function defined in [24, Eq. (9.301)].

# References

[1] J. N. Laneman, D. N. C. Tse, and G. W. Wornell, "Cooperative diversity in wireless networks: Efficient protocols and outage behavior," *IEEE Transactions on Information Theory*, vol. 50, no. 11, pp. 3062–3080, Dec. 2004.

[2] C. E. Shannon, "Two-way communication channels," in *Proc. 4th Berkeley Symp. Math. Stat. Prob*, 1961.

[3] B. Rankov and A. Wittneben, "Spectral efficient signaling for half-duplex relay channels," in *Proc. IEEE 39th Asilomar Conference on Signals, Systems, and Computers 2005 (ASILOMAR 2005), Pacific Grove, CA, USA*, Nov. 2005, pp. 1066–1071.

[4] P. Popovski and H. Yomo, "Wireless network coding by amplify-and-forward for bi-directional traffic flows," *IEEE Communications Letters*, vol. 11, no. 1, pp. 16–18, Jan. 2007.

[5] J. S. Kim, P. Mitran, and V. Tarokh, "Performance bounds for bidirectional coded cooperation protocols," *IEEE Transactions on Information Theory*, vol. 54, no. 11, pp. 5235–5241, Nov. 2008.

[6] A. Bletsas, A. Khisti, D. Reed, and A. Lippman, "A simple cooperative diversity method based on network path selection," *IEEE Journal on Selected Areas in Communications*, vol. 24, no. 3, pp. 659–672, Mar. 2006.

[7] Y. Zhao, R. Adve, and T.-J. Lim, "Improving amplify-and-forward relay networks: optimal power allocation versus selection," *IEEE Transactions on Wireless Communications*, vol. 6, no. 8, pp. 3114–3123, Aug. 2007.

[8] A. Ibrahim, A. K. Sadek, W. Su, and K. J. R. Liu, "Cooperative communications with relay-selection: when to cooperate and whom to cooperate with?" *IEEE Transactions on Wireless Communications*, vol. 7, no. 7, pp. 2814–2827, Jul. 2008.

[9] I. Krikidis, "Relay selection for two-way relay channels with MABC DF: A diversity perspective," *IEEE Transactions on Vehicular Technology*, vol. 59, no. 9, pp. 4620–4628, Nov. 2010.

[10] Q. F. Zhou, Y. Li, F. C. Lau, and B. Vucetic, "Decode-and-forward two-way relaying with network coding and opportunistic relay selection," *IEEE Transactions on Communications*, vol. 58, no. 11, pp. 3070–3076, Nov. 2010.

[11] Y. Li, R. H. Louie, and B. Vucetic, "Relay selection with network coding in two-way relay channels," *IEEE Transactions on Vehicular Technology*, vol. 59, no. 9, pp. 4489–4499, Nov. 2010.

[12] Y. Jing, "A relay selection scheme for two-way amplify-and-forward relay networks," in *Proc. IEEE International Conference on Wireless Communications and Signal Processing (WCSP 2009), Nanjing, China*, Nov. 2009.

[13] S.Atapattu, Y. Jing, H. Jiang, and C. Tellambura, "Opportunistic relaying in two-way networks," in *Proc. 5th IEEE International ICST Conference on Communications and Networking (CHINACOM 2010), Beijing, China*, Aug. 2010.

[14] L. Song, "Relay selection for two-way relaying with amplify-and-forward protocols," *IEEE Transactions on Vehicular Technology*, vol. 60, no. 4, pp. 1954–1959, May 2011.

[15] S. Atapattu, Y. Jing, H. Jiang, and C. Tellambura, "Relay selection schemes and performance analysis approximations for two-way networks," *IEEE Transactions on Communications*, vol. 61, no. 3, pp. 987–998, Mar. 2013.

[16] J. L. Vicario, A. Bel, J. A. Lopez-Salcedo, and G. Seco, "Opportunistic relay selection with outdated CSI: Outage probability and diversity analysis," *IEEE Transactions on Wireless Communications*, vol. 8, no. 6, pp. 2872–2876, Jun. 2009.

[17] M. Torabi, D. Haccoun, and J.-F. Frigon, "Impact of outdated relay selection on the capacity of AF opportunistic relaying

systems with adaptive transmission over non-identically distributed links," *IEEE Transactions on Wireless Communications*, vol. 10, no. 11, pp. 3626–3631, Nov. 2011.

[18] M. Seyfi, S. Muhaidat, and J. Liang, "Performance analysis of relay selection with feedback delay and channel estimation errors," *IEEE Signal Processing Letters*, vol. 18, no. 1, pp. 67–70, Jan. 2011.

[19] M. Soysa, H. A. Suraweera, C. Tellambura, and H. K. Garg, "Partial and opportunistic relay selection with outdated channel estimates," *IEEE Transactions on Communications*, vol. 60, no. 3, pp. 840–850, Mar. 2012.

[20] D. S. Michalopoulos, H. A. Suraweera, G. K. Karagiannidis, and R. Schober, "Amplify-and-forward relay selection with outdated channel estimates," *IEEE Transactions on Communications*, vol. 60, no. 5, pp. 1278–1290, May 2012.

[21] L. Fan, X. Lei, R. Hu, and W. Seah, "Outdated relay selection in two-way relay network," *IEEE Transactions on Vehicular Technology*, To appear. 2013.

[22] H. Cui, R. Zhang, L. Song, and B. Jiao, "Performance analysis of bidirectional relay selection with imperfect channel state information," *http://arxiv.org/abs/1112.2374*, Dec. 2011.

[23] M. Hasna and M. Alouini, "A performance study of dual-hop transmissions with fixed gain relays," *IEEE Transactions on Communications*, vol. 3, no. 6, pp. 1963–1968, Nov. 2004.

[24] I. S. Gradshteyn and I. M. Ryzhik, *Table of Integrals, Series, and Products*, 7th ed. Academic Press, 2007.

[25] F. Yilmaz, A. Yilmaz, M.-S. Alouini, and O. Kucur, "Transmit antenna selection based on shadowing side information," in *Proc. IEEE Vehicular Technology Conference (VTC 2011 Spring), Budapest, Hungary*, May, 2011.

[26] Y. Chen and C. Tellambura, "Distribution functions of selection combiner output in equally correlated Rayleigh, Rician, and Nakagami-m fading channel," *IEEE Transactions on Communications*, vol. 52, no. 11, pp. 1948–1956, Nov. 2004.

[27] Wolfram Functions, *http://functions.wolfram.com/*, 2012.

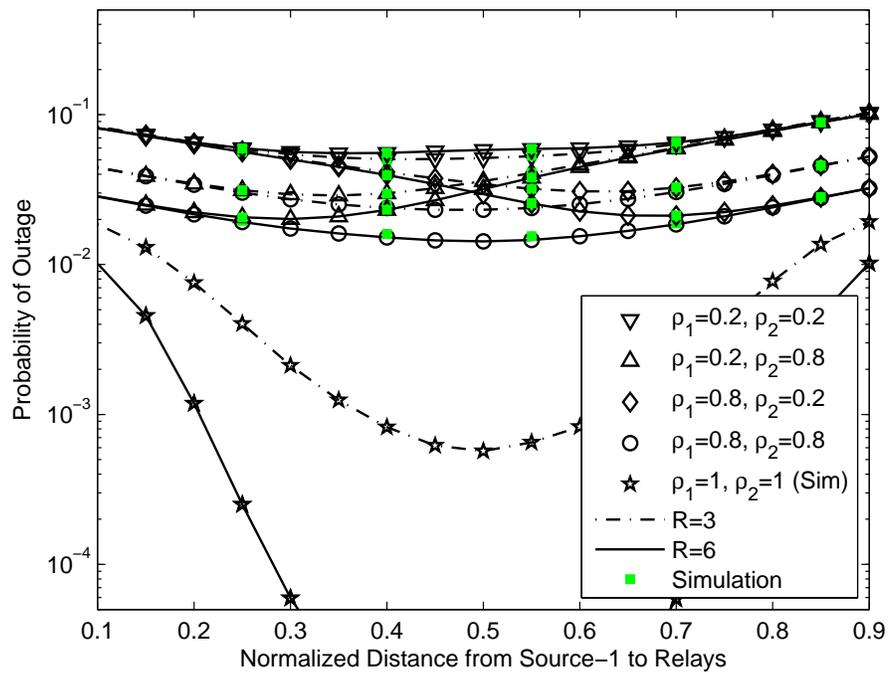

Fig. 1.   Probability of outage performance as a function of relay location where $\eta_1 = 15$ dB and $\upsilon = 3$.

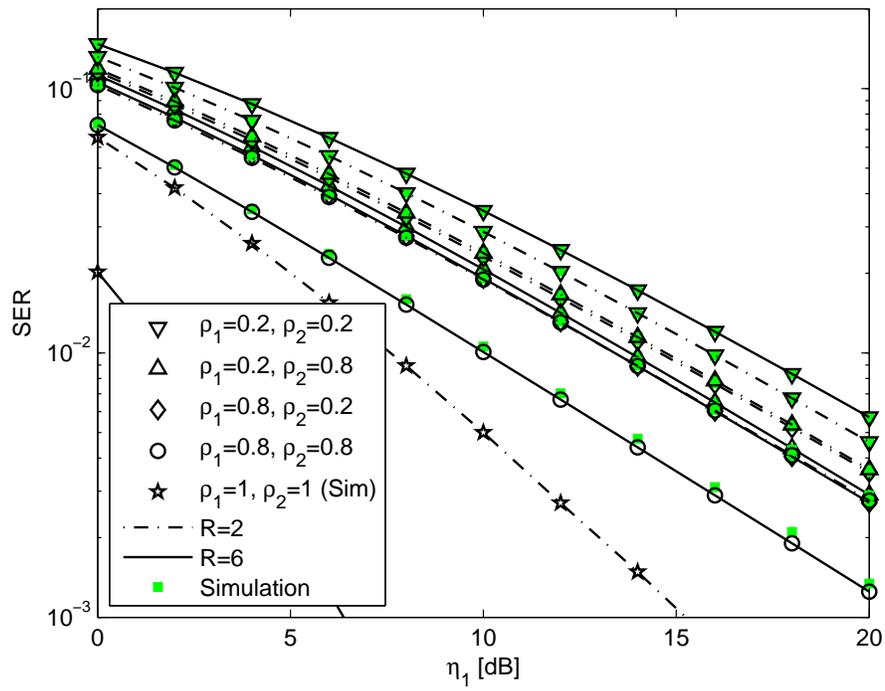

Fig. 2. Symbol error rate performance of BPSK modulation where $d_1 = 0.5$ and $\upsilon = 3$.